\documentclass[rnote]{aa} % for the research notes
\usepackage{graphicx,subfigure}
\usepackage{lscape}
\usepackage{txfonts}
\usepackage{epsfig,color}

\def\4U{4U\thinspace1705-44}
\begin{document}

%   \thesaurus{06     % A&A Section 6: Form. struct. and evolut. of stars
%              (03.11.1;  % Cosmogony,
%               16.06.1;  % Planets and satellites: general,
%              19.06.1;  % Solar system: general,
%               19.37.1;  % Stars: formation of,
%              19.53.1;  % Stars: oscillations of,
%               19.63.1)} % Stars: structure of.
%
\title{Broadband observations  of the X-ray burster \4U with BeppoSAX}

 \author{S.~Piraino \inst{1,2} \and
   A.~Santangelo\inst{2}  \and B.~M\"uck \inst{2}  \and  P.~Kaaret\inst{3}  \and T.~Di Salvo\inst{4}   \and A.~D' A\`i \inst{1} \and R.~Iaria \inst{4}  \and E.~Egron \inst{5} }

%   \offprints{S.~Piraino}

   \institute{ INAF-IASF Palermo,  Via Ugo La Malfa 153, 90146 Palermo, Italy,  \email{Santina.Piraino@ifc.inaf.it}
          \and
           IAAT University of Tuebingen, Sand 1, 72076 Tuebingen, Germany
           \and
           Department of Physics and Astronomy, University of Iowa, Iowa City, IA 52242, USA
           \and
           Dipartimento di Fisica e Chimica, University of Palermo,
via Archirafi 36 - 90123 Palermo, Italy
           \and
           INAF - Osservatorio Astronomico di Cagliari, Loc. Poggio dei Pini, Strada 54, 09012 Capoterra (CA), Italy}
   \date{Received ; accepted }
\abstract
% context
{\4U is one of the most-studied type I X-ray burster and Atoll sources. This source represents  a perfect candidate to test different models proposed to self-consistently track the physical changes occurring between    different spectral states because it shows clear spectral state transitions.}
%aims
{The broadband coverage, the sensitivity and energy resolution  of the
BeppoSAX satellite offers the opportunity to disentangle the  components that form the total X-ray spectrum and to study their changes according to the spectral
state.}
%methods
{Using two BeppoSAX observations carried out in August and October 2000, respectively, for a
total effective exposure time of $\sim 100$ ks, we study the spectral evolution of the
source from a soft to  hard state. Energy spectra are selected according to the source
position in the color-color diagram (CCD)}
%results
{We  succeeded in modeling the spectra of the source using a physical self-consistent scenario for both the island and banana branches (the double Comptonization scenario). The  components observed are the soft Comptonization and hard Comptonization, the blackbody, and a reflection component with a broad iron line.  When the source moves from the banana state to the island state, the parameters of the two Comptonization components change significantly and  the blackbody component becomes too weak to be detected.}
%conclusion
{We interpret the soft Comptonization component as emission from   the
hot plasma surrounding the neutron star, hard Comptonization  as emission from the disk region, and the blackbody component as emission from the
inner accretion disk. The broad feature in the iron line region is compatible with reflection from the inner accretion disk.}


\authorrunning{Piraino et ~al.\ 2016}
\titlerunning{BeppoSAX observation of \4U}

\maketitle

\section{Introduction}

Hasinger \& van der Klis (1989) divided low mass X-ray binaries (LMXBs) in two
groups, called Z and atoll sources after the patterns these sources
trace out in the X-ray color-color diagram (CCD). The CCD of the Z sources
displays a Z-like track, whereas the CCD of atoll sources show a C-like track
(M\'endez 1999) where two branches can be identified. These are the island and  banana
states. The island branch is characterized by lower count rates, much less variability in the CCD, and stronger band-limited
noise than the
banana state.

It has been shown, however, that when sampling the source intensity states of the
atoll sources for a long enough time, the shape of their CCD tends to resemble those
of Z sources (Gierl\'{i}nski \& Done 2002; Muno, Remillard \& Chakrabarti
2002). The transient LMXB XTE J1701-462 was the first source to show
transitions between all the branches of Z and atoll sources within an outburst lasting 
a few months  (Lin, Remillard  \& Homan 2009). Also, the transient X-ray pulsar in 
Terzan 5, IGR J17480-2446, showed atoll and Z-source characteristics during an outburst (Altarimano et al. 2010).

The bright neutron star LMXB \4U  was classified  by Hasinger \& van der
Klis 1989, as an atoll source. However, long-term monitoring of this source
showed a pattern similar to a  complete Z track in the CCD (Gierl\'{i}nski \&
Done 2002, Muno et ~al.\ 2002, Barret \& Olive 2002). 

\4U is an X-ray burster (Sztajno et~al.\ 1985; Langmeier et~al.\ 1987), whose bursting
activity and frequency depends on persistent emission.  The source is characterized by variability between low and high intensity states. 
During the low intensity state, when type I X--ray bursts are most frequent, the spectrum is hard (Langmeier et~al.\
1987). Barret \& Olive (2002), monitoring \4U with the Rossi X-Ray Timing Explorer, observed a spectral state transition
soft-hard-soft correlated with the intensity, from a 0.1-200 keV luminosity of 6.9 to 31$\times$ 10$^{36}$~erg~s$^{-1}$, of a distance to the source of 7.4 kpc (Haberl \& Titarchuk 1995).
These authors interpreted the observed spectral evolution within a scenario of a truncated
accretion disk of varying radius and an inner flow merging smoothly with the
neutron star boundary layer (Barret et~al.\ 2000, Done 2002). 
 Barret \& Olive (2002) successfully modeled the spectra with a blackbody plus a
dominating Comptonized emission (modeled in XSPEC with {\tt compTT}). An
iron line whose energy was fixed at 6.4 keV was also detected. 
The soft component (blackbody) was interpreted as originating from the disk at higher luminosity and
from the neutron star surface at lower luminosity. The Comptonized component
was associated with the hot  inner flow. Barret \& Olive 2002 showed that the truncation
radius is not just set by the instantaneous $\dot{M}$, as observations with the
same bolometric luminosity have very different spectral and timing properties.

Di Salvo et al. (2005) observed \4U in the energy range 0.3-10 keV with  Chandra, during a
soft state at a luminosity of 33$\times$ 10$^{36}$~erg~s$^{-1}$, and modeled the spectral continuum with a soft Comptonization
model {\tt compTT} (electron temperature kT$_e$ $\sim$  2.3 keV and optical depth $\tau
\sim$ 18 for a spherical geometry). 
The iron line K$_\alpha$ at 6.5 keV was found to be intrinsically broad and compatible with a reflection from a cold accretion disk with R$_{in}$ $\sim$
15 km or with a Compton broadening in the external parts of a $\sim$2 keV corona.

A Newton-XMM  observation, performed when the source was in a soft state, at $\sim$~1.0$\times $10$^{38}$~erg~s$^{-1}$, showed that the reflection scenario is the favored solution for explaining the broad emission features (Di Salvo et~ al.\ 2009, D' A\`i et~al.\  2010).
The broad line of  \4U, with others lines detected with Newton-XMM, is also discussed in Ng et~al.\  2010 who point out the potential role of the pileup in the broadening of the line. 
However, recent developments of the subject can be found in Di Salvo et~al.\ 2015, Cackett \& Miller 2013, Sanna et~al.\ 2013,  Egron et~al.\ 2013, Piraino et~al.\ 2012,  Cackett et~al.\ 2012, and Miller et~al.\ 2010. 

We report a detailed spectral and timing analysis of two BeppoSAX observations of the source performed on August and October 2000. 
The same data have already been partially used by Fiocchi et al. 2007, Lin et al. 2010, Cackett et al. 2012, and Egron et al. 2013.
In fact, our analysis alone uses all BeppoSAX NFIs for both observations. 
Moreover, in our contribution we model the X-ray broadband energy (0.3--200~keV) spectrum of
\4U obtained for four different CCD positions (three in the banana state
and one in the island state). In particular, we try to fit the both hard and soft spectra with
the same model components to highlight the evolution of the
spectral parameters along the CCD or hardness intensity
diagram. This  attempt is similar to the approach of Seifina et al.
(2015) based on BeppoSAX and RXTE spectra. 
We also present, for the first time, the  BeppoSAX Fourier power density spectrum of the source
variability in the overall banana and island states. 
We present the observations, results of the timing, and spectral analysis in $\S$ 2, 3, and
4, respectively, and discuss the results in $\S$ 5.

\section{Observations}

Two BeppoSAX observations of \4U were performed August 20--21 2000 (OBS1) and October
3--4 2000 (OBS2), respectively, for a total of 100~ks of on-source observing time. 

We obtained spectra from the four BeppoSAX Narrow Field Instruments (NFIs) in
energy bands selected to provide good signal to noise for this source: the Low
Energy Concentrator Spectrometer (LECS; 0.1--4~keV; \cite{Parmar97}), the
Medium Energy Concentrator Spectrometer (MECS; 2--10~keV; Boella et al.
\cite{Boella97}), the High Pressure Proportional Gas Scintillation Counter
(HPGSPC; 4--34~keV; \cite{Manzo97}), and the Phoswich Detection System (PDS;
15--200~keV; \cite{Frontera97}). We extracted the LECS and MECS data in circular
regions centered on the source position using radii of 8' and 4', respectively,
containing 95\% of the source flux. We used the same circular regions in the blank field
data  for background subtraction. As the source is centered along
the Galactic bulge, we scaled the blank field background at the source position to
the mean level of the background around the source  using
a factor 2.7 for LECS and 2.64 for MECS. Spectra accumulated from dark Earth
data and during off--source intervals were used for background subtraction
in the HPGSPC and PDS respectively.

We rebinned the LECS and MECS  spectra to sample the
instrument resolution with the same number of channels at all
energy, and we grouped the HPGSPC and PDS spectra using a
logarithmic grid.  A normalization factor has been included to
account for the mismatch in the BeppoSAX instruments absolute
flux calibration. The fit values of relative normalization are
in good agreement with values typically observed (Fiore et~al.\
1999).

\section{Timing  analysis}
\begin{figure}
\centering
\hbox{\hspace{0cm}
\includegraphics[width=8.7cm]{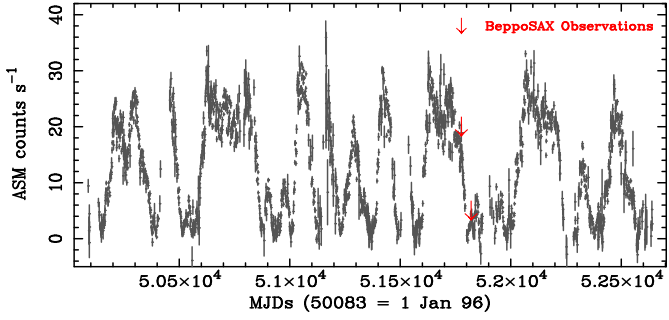} }  
\parbox[b]{8.7cm}{\caption[] {Long-term (Jan 1996 -- Dic 2002) RXTE/ASM 2-12 keV light curve
(gray). The data were retrieved from HEASARC public data base. The time of  BeppoSAX
pointed observations (MDJs=51766/7  and 51820/1)   are indicated with arrows. During two
observation the source intensity changes from $\sim$ 300 mCrab to $\sim$ 30 mCrab.
Overall the source intensity can reach $\sim$ 500 mCrab.} \label{fig1}}
\end{figure}

Fig.~\ref{fig1} shows  the RXTE-All Sky Monitor (ASM) light curve (over six
years) of \4U, where the high long-term X-ray intensity variability is clearly visible. The
data of the BeppoSAX pointed observations are indicated with arrows. Observation OBS1 (MJDs=51766/7) 
occurred at the beginning of a clear transition from a
high to a low intensity state, while during OBS2 (MDJs=51820/1) the source was in
a low intensity state.

\begin{figure}
 \centering
 \hbox{\hspace{0cm}
\includegraphics[width=4.3cm]{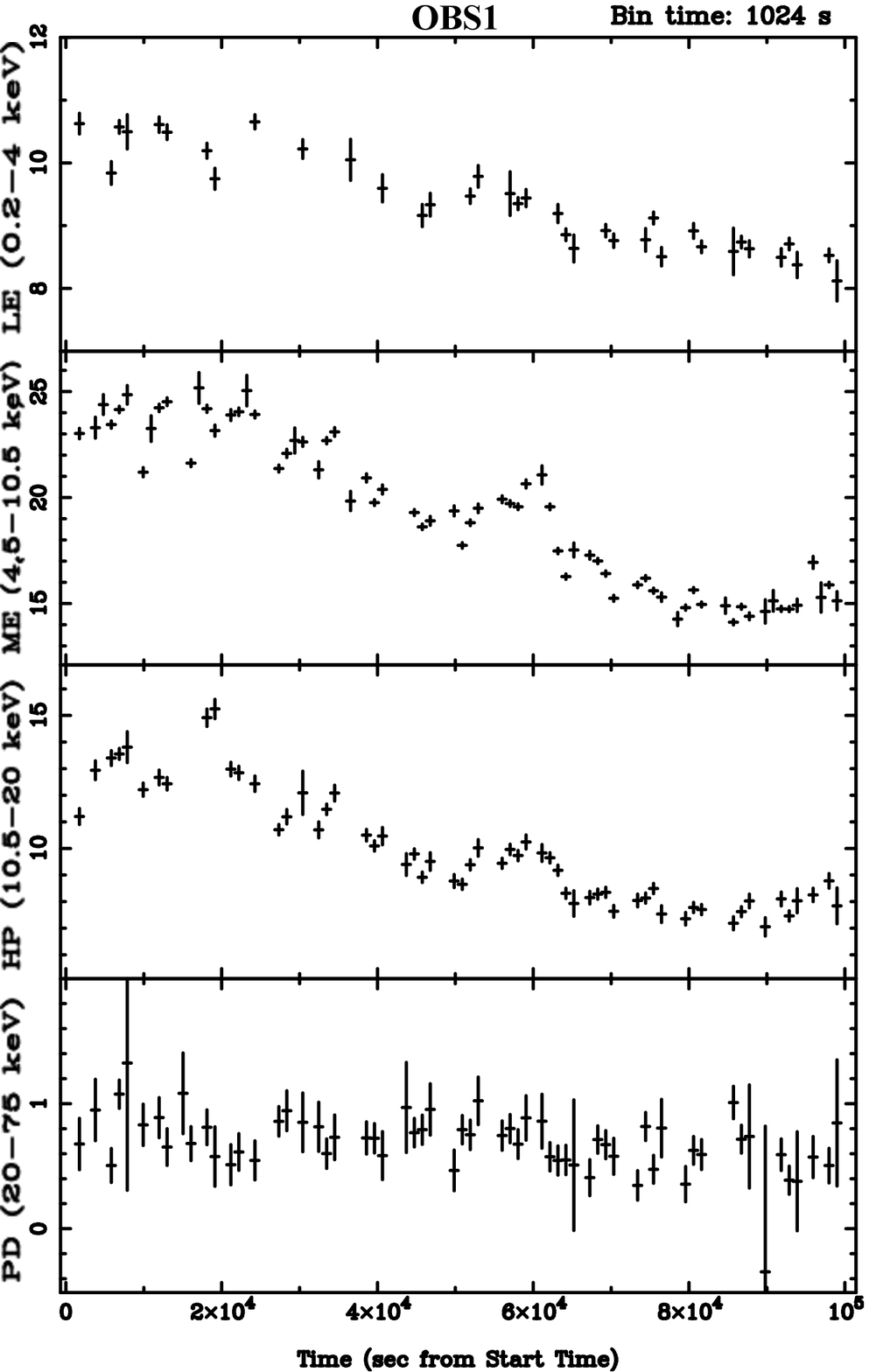}  
\includegraphics[width=4.5cm]{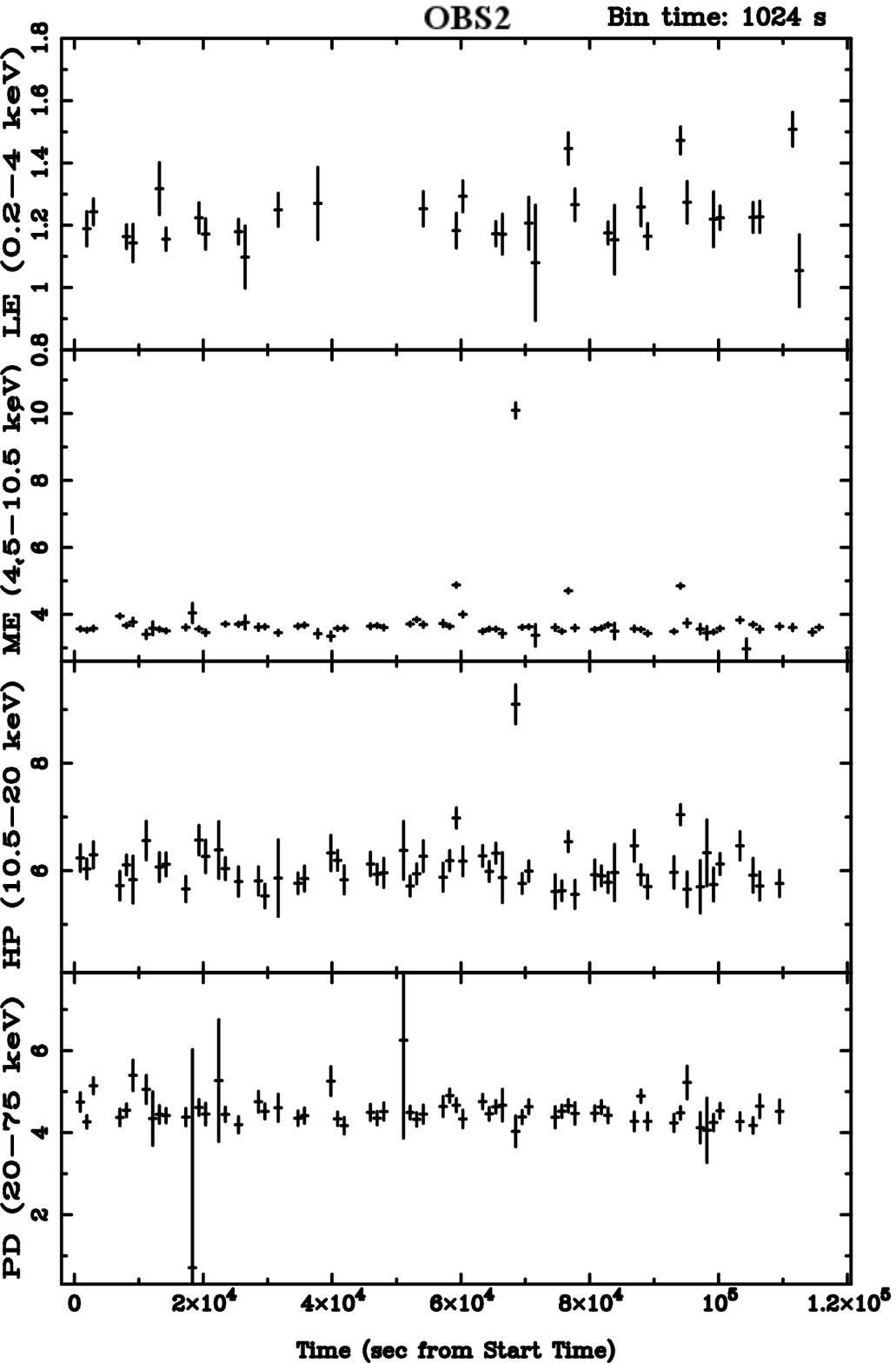}}
\parbox[b]{8.7cm}{\ \caption{BeppoSAX NFIs light curves during OBS1 (left) and OBS2 (right). The excess
rate in the first tree panels of OBS2  are from X-ray bursts.}\label{lcurve}}
\end{figure}

BeppoSAX light curves obtained from OBS1 and OBS2 are shown, in different
energy range of NFIs, in Fig.~\ref{lcurve}.
A bin size of 1024~s was used. During  OBS1 the overall intensity of the source
changed up to a factor $\sim$2 in the range 4.5--30~keV. Six X--ray bursts occurred during the OBS2 observation and were excluded in the subsequent analysis.  

\begin{figure} \vspace{0cm}
\hbox{\hspace{0cm}
\includegraphics[width=4.3cm]{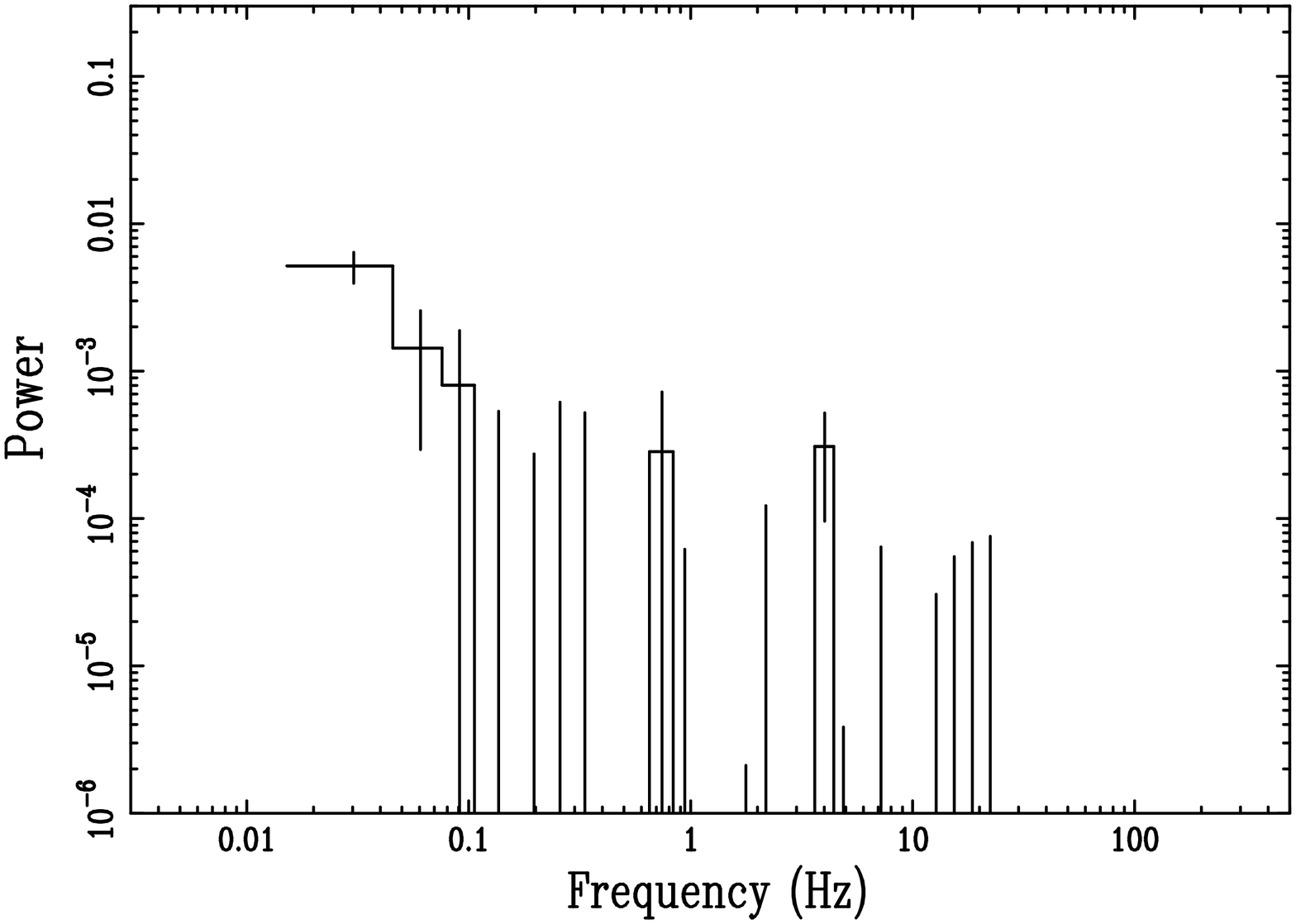}
%\hspace{-0.4cm}
\includegraphics[width=4.3cm]{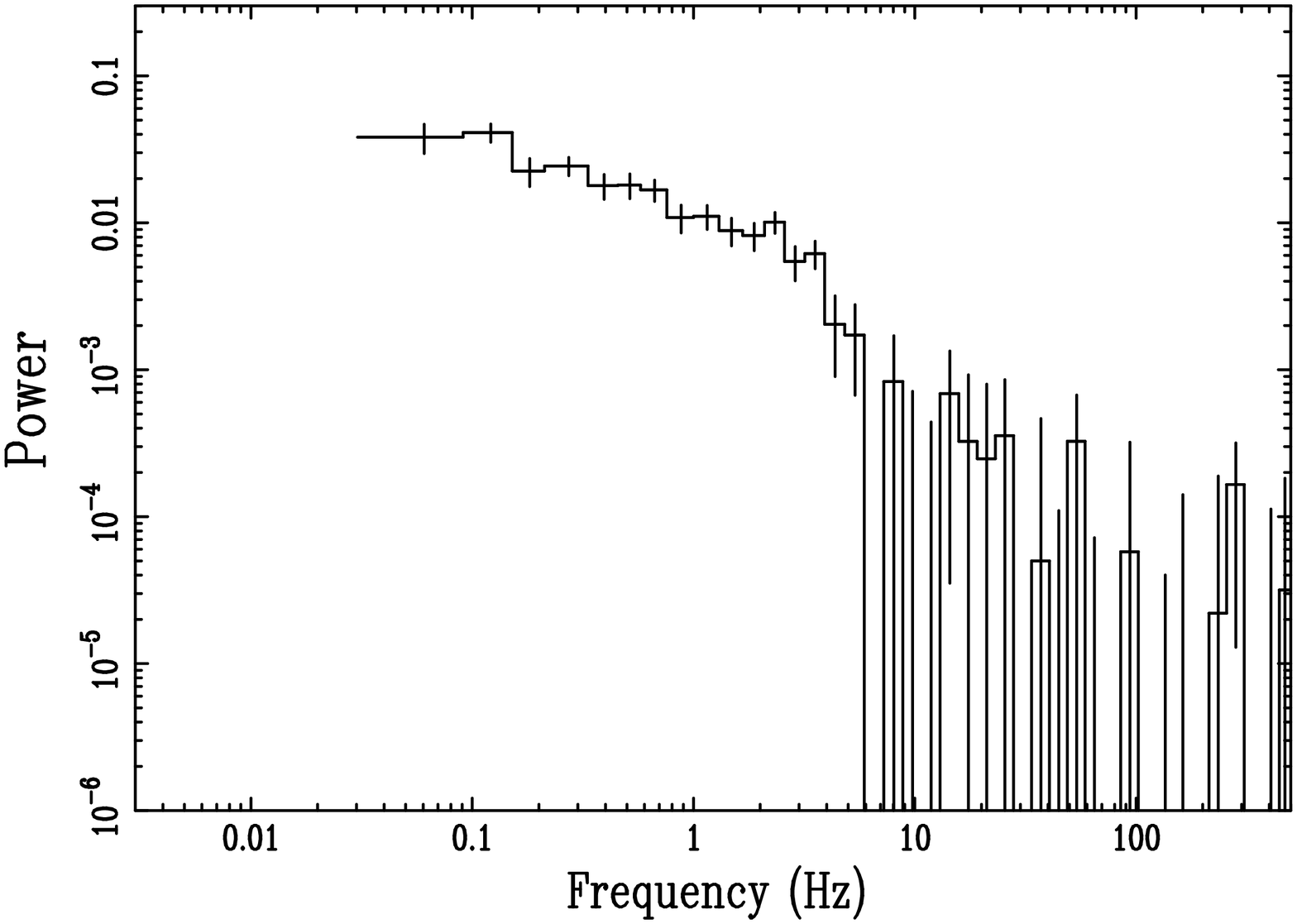}} \vspace{0cm} {\caption[]{Left panel: PSD in
the banana state. Right panel:  PSD in the island state} \label{psd}}
\end{figure}

Fig.~\ref{psd} shows the power spectral density (PSD) of the OBS1 and OBS2 observations, computed using
the MECS data, background (Poisson counting noise
level) subtracted, rebinned using a logarithmic scale, and
expressed in terms of fractional rms amplitude. In the lower intensity OBS2, the source shows strong band-limited noise and no very low frequency
noise, therefore suggesting that the source was in the island state.
The observed PSD is typical for these systems  characterized by a Lorenzian shape with a break frequency around 0.2 Hz. The PDS  for OBS1 is featureless, indicating that the variability during this observation was low.

\begin{figure}
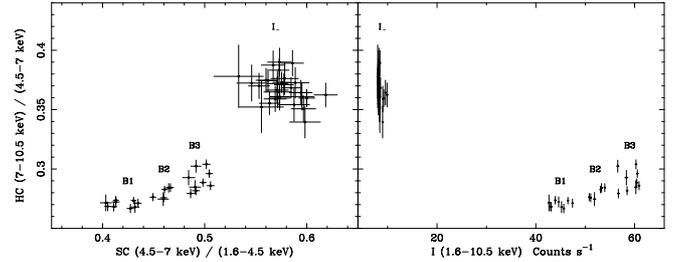
 \vspace{0cm}
\centering
\hbox{\hspace{0cm} \psfig{file=cc_2016.ps,width=3.4cm,angle=270.0}
\hspace{-0.1cm}\psfig{file=hid_2016.ps,width=3.4cm,angle=270.0}}
\vspace{0cm}
\parbox[b]{8.7cm}{\caption[]{Left panel: color-color  diagram.
Right panel: hardness intensity diagram; the banana and island branches are clearly
visible. Energy spectra were extracted in region B1, B2, B3, and I.} \label{ccd}}
\end{figure}

 Fig.~\ref{ccd} shows the CCD and the hard color versus intensity diagram (HID) of 4U 1705-44, where
the hard color (HC) is the ratio of the 7--10.5~keV to 4.5--7~keV MECS counts rates
and the soft color (SC) is the ratio of the 4.5--7~keV to 1--4.5~keV MECS counts
rates. The intensity is the count rate in the 1--10.5~keV energy range.  Each point corresponds to a time-averaged interval of 4096 s.
From the CCD and HID diagrams, we conclude that, during OBS1, the source resided in the lower branch of the Z track
reported by Gierlinski \& Done (2002), Muno et al. (2002), and Barret \& Olive
(2002). In OBS2, the source is confirmed in the island state, even if the colors are unusual, in that they are very
similar to those of the very high part of the banana state. This peculiarity was also observed based on EXOSAT data of \4U (Berger \& van der Klis 1998).

In order to investigate the spectral evolution of \4U along the CCD, we extracted the energy spectra of each NFI over 
four different SC intervals defined by boundaries 0.40-0.44, 0.45-0.47, 0.48-0.51, and 0.53-0.62. We identified three data  regions, labeled B1, B2, and  B3 in Fig.~\ref{ccd} in the banana state (OBS1), while I is the island state and covers the whole OBS2.

\section{Spectral analysis}

The averaged spectrum of the source for OBS1, that is in the banana branch, was analyzed in Piraino et al. 2007.
It is well fitted by a blackbody ($kT_{bb}\simeq0.56$~keV),
contributing $\sim$14\% of the observed 0.1--200~keV flux, plus a Comptonized
component modeled with {\tt compTT} (seed photon temperature of $\sim$1.1~keV,
electron temperature $\sim$2.7~keV, and optical depth $\sim$11). A hard
power-law  component (photon index $\sim$2.9), contributing about 11\% of the
0.1--200~keV source flux, was significantly detected. A very broad K$_{\alpha}$  line at 6.4--6.7~keV (see the above mentioned paper for details) was also reported.
The 0.1--200~keV source   luminosity is  $\sim$ 10$^{38}$~erg~s$^{-1}$ assuming a distance to the source of 7.4~kpc (Haberl \& Titarchuk
1995).

\begin{figure}[ht]
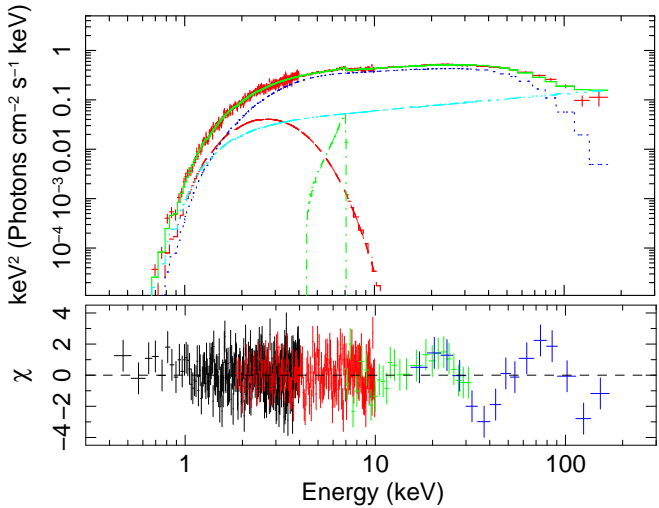
 
\begin{minipage}[t]{0.4\linewidth}
\centering
\vbox{\psfig{file=eeuf_h1model1.ps,height=8.5cm,angle=270.0}\vspace{-0.1cm}
\psfig{file=delchi_h1model1.ps,height=8.5cm,angle=270.0}}
\parbox[b]{8.7cm}{\caption[]{Upper panel: unfolded energy spectrum with Piraino et al. 2007 model  and
single components in region I. Botton panel: residuals.} \label{h1model1}}
\label{fig:himodel1}
\end{minipage}
\end{figure}

To investigate the spectral evolution of \4U on the CCD, we 
fit the broadband spectra extracted from four CCD regions (see Fig.~\ref{ccd}) with the model described above.
The model adequately fits  the data of B1, B2, and B3. However, the spectral continuum of  the island state I, fitted with the above mentioned model, shows marked residuals at higher energies (see Fig.~\ref{h1model1}). 

\begin{figure}[ht] 
\begin{minipage}[t]{0.4\linewidth}
\centering
\vbox{ \psfig{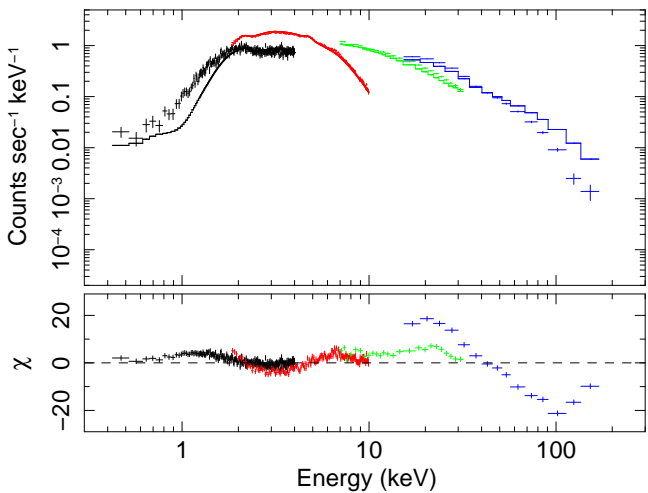}} \vspace{0cm}
\parbox[b]{8.7cm}{\caption[]{Upper panel: folded spectra and absorbed power-law model. Botton panel: residuals.} \label{h1modelpow}}
\end{minipage}
\end{figure}

Unfortunately, none the two-component  models proposed  for the continuum of LMXB, including the western
(White, Stella \& Parmar 1988), eastern (Mitsuda et al. \ 1989), Birmingham (Church \& Balucinska-Church 1995), or hybrid (Lin et al. 2007),  can describe the broadband I spectrum, leaving significant residuals in the highest band. We therefore propose a more complex model to take the origin of this broad residual pattern into account.

To better highlight  the structure of the spectrum in Fig.~\ref{h1modelpow}, we show the result of a fit with a simple absorbed
power-law  (the $\chi^{2}$/d.o.f was 4801/491). In the residuals we can clearly observe  a soft excess, an iron line, and a broad hump over 20-40 keV. That may suggest reflection from a disk.  Similar features  have been observed in black hole binaries (e.g. Di Salvo, Done, Zychi et~al.\ 2001;  Makishima et~al.\ 2008).

To model the spectrum we firstly  used  the Poutanen \& Svensson (1966) Comptonization model, 
({\tt compPS} in XSPEC), which as a built-in function to allow part of
the Compton-produced photons to be reflected by cold matter (disk). We thus
fitted the I spectrum with the model  {\tt wabs*(diskline+compPS)} \textbf{(Model1)}, assuming a spherical geometry for the Compton cloud. We activated
the reflection option, assuming  a solar abundance for the cold disk matter. We can successfully model the whole (0.3 --200 keV) spectrum  ($\chi^{2}/\nu$ = 1.03, $\nu$=484), obtaining a subtended solid angle of
$\Omega/2\pi \sim 0.3$ (See first column in  Table 1).  

Unfortunately, \textbf{Model1}  cannot be applied to the B1, B2, or B3 spectra because in the banana state the electron cloud temperature is $\sim$3~keV (Piraino et al. 2007, Di Salvo et al. 2009, Egron et al. 2013), while the compPS component was considered for electron cloud temperatures T$_e > $10 keV.
We  also tried to fit the  island state spectrum I using the alternative model {\tt wabs*(diskline+compTT+compPS)} (\textbf{Model2}), containing a soft Comptonization component ({\tt compTT}) with spherical geometry, plus a second hard Comptonization component ({\tt compPS}) with disk geometry. In \textbf{Model2} the soft (\emph{secondary}) Comptonization is thought to originate in the inner region close to the neutron star, where kT$_i$ and kT$_e$ is equal to 0.8 and 6.7 keV and $\tau \sim 8$, while the hard (\emph{main}) Comptonization
{\tt compPS} comes from a higher temperature region above the disk, which is characterized by kT$_e$=22 keV,  kT$_i$=0.11 keV and $\tau$=2.9. 
The reflection component  subtends an angle of $\Omega/2\pi \sim 0.3$ (see  second column in Table 1).
\begin{table*}

\caption[]{Best-fit parameters of \textbf{Model1} und \textbf{Model2}}
%Results of the fit of \4U I,B1,B2,B3 spectra in the energy band 0.3--200~keV.

\begin{center}
\label{tab1}
\begin{tabular}{l|c||c|c|c|c}
\hline \hline
&  \multicolumn{1}{c}{\textbf{Model1}}   & \multicolumn{4}{c}{\textbf{Model2}}  \\

&  \multicolumn{1}{c}{\tt wabs(compPS+diskline) }  & \multicolumn{4}{c}{\tt wabs(bb+compTT+compPS+diskline)} \\

  &\multicolumn{1}{c} {}  & \multicolumn{4}{c}    {}     \\
  \hline
 CDI Region  & {\tt I} & {\tt I}&  {\tt B1}  &  {\tt B2}  &  {\tt B3} \\
 \hline

$N_{\rm H}  \rm (\times 10^{22}\;cm^{-2})$  & $1.18  \pm 0.05$ &  $1.59  \pm 0.03$ &  $  1.81\pm0.03 $   &$ 1.76 \pm0.04 $& $ 1.78 \pm0.04 $         \\

$k T_{\rm bb}$ (keV)              &** &  **       & $ 0.57 \pm0.03 $    &$ 0.58 \pm0.02 $&     $ 0.58 \pm0.02 $           \\

$N_{\rm bb}\times 10^{-3}$       &**& **  &  $  19\pm2 $         &$  24\pm1 $&  $  24\pm1 $            \\

$k T_{\rm W}^{\rm cTT}$ (keV) &&             $0.78 \pm 0.02$      &$ 1.02 \pm0.07 $  &$  1.15\pm0.06 $& $ 1.16 \pm0.06 $       \\

$k T_{\rm e}^{\rm cTT}$ (keV) &&             $6.7 \pm 0.3$     &$  2.65\pm0.08 $ &$ 2.78 \pm0.09 $&    $ 2.71 \pm0.07 $     \\

$\tau^{\rm cTT}$                         &&             $7.9 \pm 0.3$  &$ 11.7 \pm0.6 $&$ 10.4 \pm0.7 $   & $ 11.3 \pm0.7 $         \\

$N_c^{\rm cTT} (\times 10^{-2}$) &&             $1.7 \pm 0.1$    &$  30\pm3 $&$ 34 \pm2 $&    $ 42 \pm3 $       \\

$k T_{\rm e}^{\rm }$ (keV)      & $27 \pm 1$ &   $21.6 \pm 0.2$         &        $31\pm5$           &$35^{+5}_{-12} $& $32^{+7}_{-4} $         \\                        
                                        
$k T_{\rm i}^{\rm }$ (keV)      &        $0.74 \pm 0.02$ &       $0.11 \pm 0.08$   &        $0.48\pm0.04 $         &$  0.47\pm0.04 $&$ 0.52 \pm0.02 $                \\
                                                                                
$\tau^{\rm cPS}$        &        $2.8 \pm 0.1 $         & $2.94 \pm 0.03 $ &$ 0.7 \pm0.1 $       &$  0.6\pm0.2 $& $0.6^{+0.4}_{-0.3} $              \\
                                        
%$Y^{\rm cPS}$          &        $3.7 \pm 0.4 $         &               &&                \\

$Rel_{refl}$    &        $0.24\pm0.04$ &         $0.27\pm0.05$  &       $  2.4\pm1 $          &$  3.0\pm2 $& $  4\pm2 $          \\

$\xi$ (erg cm s$^{-1}$)         &        $393^{+124}_{-167}$&    $1470 \pm 600$    & $ 3800^{+1500}_{-700}  $      &$2600 ^{+1300}_{-1000}  $&  $ 2800^{+1500}_{-1000} $       \\

$N_c^{\rm cPS} (\times 10^{3}$)  &       $201\pm20 $    &        $30.7\pm5 $       &       $ 1.0^{+0.5}_{-0.2}  $&$  0.7^{+0.9}_{-0.3} $&$  0.5^{+0.3}_{-0.2} $      \\
$R_{\rm i}^{\rm cPS}$ (km)      &        $ 11 \pm  1$  &&&&      \\

$E_{\rm Fe}^{\rm diskline}$ (keV)&  $6.44\pm 0.09$ &  $6.47\pm 0.08$ &  $6.43\pm0.07 $    &$  6.48\pm0.08 $&    $  6.42\pm0.05 $      \\

$R_{\tt in}$ (M)  &      $9^{+6}$ & $12^{+5}$ &  $7^{+5}$     &$  13^{+14} $&  $  9.5^{+4} $        \\

$Incl$ (deg) & $40.$ fix & $40.$ fix & $40.$ fix    &$40.$ fix&$40.$ fix           \\

$I_{\rm Fe}$ ($\times 10^{-3}$ ph cm$^{-2}$ s$^{-1}$)   &        $0.9 \pm 0.4   $ & $1.1 \pm 0.2  $ & $ 4.2\pm0.5   $     &$  4.2\pm0.8 $& $  6.7\pm0.9 $       \\

           &            &   &&  &                     \\

$\chi^2_{\rm red}$ (d.o.f.) &            1.03 (485)& 1.05 (482) & 1.03 (542)     &0.97(544)&1.08(412)           \\

\hline
\end{tabular}
\end{center}
{\sc Notes} \footnotesize---{\textbf{Model1}  consists of a Comptonization  component ({\tt compPS}) and a reflection with emission line ({\tt diskline}). \textbf{Model2} consists of a blackbody model ({\tt bb}) and a double Comptonization component ({\tt compTT} and {\tt compPS}  with reflection) and the  emission line. The two asterisks indicate that the {\tt bb} component is not required in the I state. When it is added to the model with $k T_{\rm bb}$ fixed to 0.1 keV the value of  $N_{\rm bb}$ is $(6 \pm7)\times 10^{-3}$; the values of the other parameters remain almost unchanged. Uncertainties are at the 90\% confidence level for a single parameter.}
\end{table*}
\begin{table*}

\caption[]{Best-fit parameters of \textbf{Model3} und \textbf{Model4} }

\begin{center}
\label{tab2}

\begin{tabular}{l|c|c|c|c||c|c|c|c}

\hline \hline

& \multicolumn{4}{c}{ \textbf{Model3}} & \multicolumn{4}{c}{\textbf{Model4}} \\

& \multicolumn{4}{c}{\tt wabs(bb+compTT+compPS+diskline)}  & \multicolumn{4}{c}{\tt wabs(compTB+compTB+diskline)}\\

& \multicolumn{4}{c}{}&\multicolumn{4}{c}{} \\

\hline
 CDI Region                                                       &  {\tt I}&  {\tt B1}  &  {\tt B2}  &  {\tt B3}   &{\tt I}       &  {\tt B1}     &  {\tt B2}  &  {\tt B3}\\
 \hline
$N_{\rm H}  \rm (\times 10^{22}\;cm^{-2})$&$1.66\pm0.04$ & $1.67\pm0.04$  &$1.67\pm0.04$     & $1.68\pm0.04$     & $1.59  \pm 0.04$& $1.54  \pm 0.05$& $1.55  \pm 0.06$& $1.56  \pm 0.05$        \\
                                                        &           &              &              &              &              &                  &              &                    \\
$k T_{\rm bb}$ (keV)                                    & $0.1 fix$        & $0.56\pm0.03$  &$0.58\pm0.03$    &$0.58\pm0.03$                &&&&           \\

$N_{\rm bb}\times 10^{-3}$                              &$10\pm10$                 & $20\pm3$         &$22\pm3$       &$23\pm3$                    &&&&           \\
                                                                         &&&  &            &   &&  &                     \\
$\alpha_{\rm c1}$                         &&&&&$0.90 \pm 0.02$&$1.4 \pm 0.1$&$1.7\pm0.2$&$1.4\pm0.2$      \\
$k T^{\rm i}_{\rm c1}$ (keV)                              &$0.78 \pm 0.02$& $1.05\pm0.07$  &$1.14\pm0.06$   &$1.16\pm0.06$    &$0.77 \pm 0.02$&$0.99 \pm 0.07$&$1.13\pm0.06$&$1.13\pm0.07$      \\

$k T^{\rm e}_{\rm c1}$ (keV)                    &$6.9 \pm 0.3$     & $2.7\pm0.1$     &$2.7\pm0.1$             &$2.7\pm0.1$         &$8.6 \pm 0.2$ &$ 3.0 \pm0.2 $&$ 3.1 \pm0.3 $&$ 2.9 \pm0.2 $    \\

$\tau_{\rm c1}$                                                 &$7.8\pm0.3$     & $11.1\pm0.8$    &$10.9\pm0.8$     &$11.4\pm0.8$                  &&&&         \\

$N_{\rm c1} \times 10^{-2}$                      &$1.6 \pm 0.1$   &$30\pm4$          &$36\pm4$           &$43\pm4$                              &$6.7 \pm 0.5$& $42 \pm 1$& $53 \pm 1$ & $62 \pm 2$     \\

$\alpha_{\rm c2}$                         &&&&&$0.6 \pm 0.1$&$0.$ fix&$0.$ fix&$0.$ fix      \\
$k T^{\rm i}_{\rm c2}$ (keV) &$0.1\pm0.001$&$0.48\pm0.01$&$0.48\pm0.01$& $0.47 \pm 0.01$                                                &$0.13\pm0.1$ &$0.52\pm0.02$&$0.54\pm0.02$&$0.54\pm0.02$              \\

$k T^{\rm e}_{\rm c2}$ (keV) &$21.1\pm0.2$&$32\pm1$    &$21\pm0.7$         &$20^{+1}$                                                          &$19\pm2$& $12^{+5}_{-3}$   & $11^{+7}_{-4}$ & $8^{+5}_{-3}$         \\                    
                                                                                
$\tau_{\rm c2}$                                         & $2.96 \pm 0.01$ &$0.9\pm0.03$       &$1.46\pm0.04$ & $1.49\pm0.04 $                  & &&   &        \\

$N_{\rm c_2}  \times 10^{3}$ & $33.4\pm0.8$   &$0.75\pm0.02$&$0.77\pm0.03$&$0.88\pm0.03$ & & & & \\
$N_{\rm c_2} \times 10^{-3}$         & & & &                                         & $1.8\pm0.2$& $18^{+1}_{-2}$& $20\pm1$& $20\pm1$          \\

$E_{\rm Fe}^{\rm diskline}$ (keV)                       &$6.49\pm 0.08$  &$6.46\pm 0.08$ &$6.5\pm 0.1$& $6.45\pm 0.09$                 &$  6.48\pm0.05 $& $  6.5{+0.1}_{-0.2} $&$  6.5\pm0.1 $& $  6.45\pm0.05 $     \\

$R_{\tt in}$ (M)                                                &$9^{+5}$         & $6^{+7}$&$6^{+9}$& $10^{+6}$                                              &$9^{+5}$ &  $6^{+10}$  &$14^{+11}$& $10^{+4}$     \\

$I_{\rm Fe}$ ($\times 10^{-3}$ ph cm$^{-2}$ s$^{-1}$) &$1.4\pm0.2$&$4.7 \pm 0.8  $ &$5.7 \pm 0.9  $ &        $7 \pm 1   $                                              &$1.3\pm0.2$&$4.8 \pm 0.2$&$4.9 \pm 0.7$&$7.5 \pm 1.0$       \\
%&&&  &            &   &&    &                  \\

                                                                         &&&  &            &   &&   &                  \\

$\chi^2_{\rm red}$ (d.o.f.)                             &1.09 (481)&1.05 (544) &0.98 (546)&1.07 (414)                                                       &1.08 (481)  &1.05 (546) &0.98 (548)&1.07 (416)         \\

\hline
\end{tabular}
\end{center}
{\sc Notes} \footnotesize---\textbf{Model3}  consists of a blackbody model and a double Comptonization component ({\tt compTT} and {\tt compPS} without reflection) and an emission line. \textbf{Model4} consist of two {\tt compTB} components and the emission line. Uncertainties are at the 90\% confidence level for a single parameter.
\end{table*}

Adding a blackbody component {\tt bb} to \textbf{Model2}, to take the direct emission from the
disk into account, we obtain a kT$_{bb}$ value of  about 0.1 keV, which is very close to that found for the input photon of {\tt compPS}. The contribution of this component to the total flux is about 10\% of the total flux. However, since the addition of a {\tt bb}  is not statistically significant for \textbf{Model2}, no final conclusion can be obtained about the flux of the direct blackbody emission originating from the disk. 
 
Using  \textbf{Model2}  on the average spectrum of the banana states,  we obtain a statistically good  fit. In this spectrum,  the {\tt bb}  component comes out to be statistically significant and the {\tt compPS}  component well describes the hard excess reported in Piraino et al. 2007. However,  as shown in the last three column of Table 1, the \textbf{Model2} used in  the three banana states  provides values of the best-fit parameters for the reflection fraction, $\Omega/2\pi>1$. As a reflection fraction of 1 corresponds an isotropic source above the disk plane, such high values suggest that the model  \textbf{Model2} is not an adequate physical model for this state.

\begin{figure}[] 
\centering
\vspace{0.5cm}
%\vbox{\psfig{file=figure_2015/luglio2015/eeuf_delchi_hard_2016.ps,height=8.5cm,angle=270.0}}
\vbox{\psfig{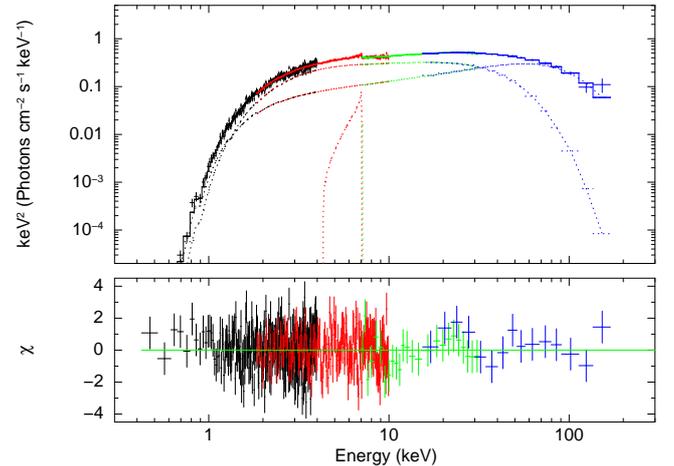}}

\vspace{0cm}
\parbox[b]{8.7cm}{\caption[]{Upper panel: unfolded energy spectrum with model \textbf{Model4} and
single components in region I. Botton panel: residuals }
\label{model4_h}}
\end{figure}

\begin{figure}[] 
\centering
\vspace{0.5cm}
%\vbox{\psfig{file=figure_2015/luglio2015/prova_spettro.ps,height=8.5cm,angle=270.0}}
\vbox{\psfig{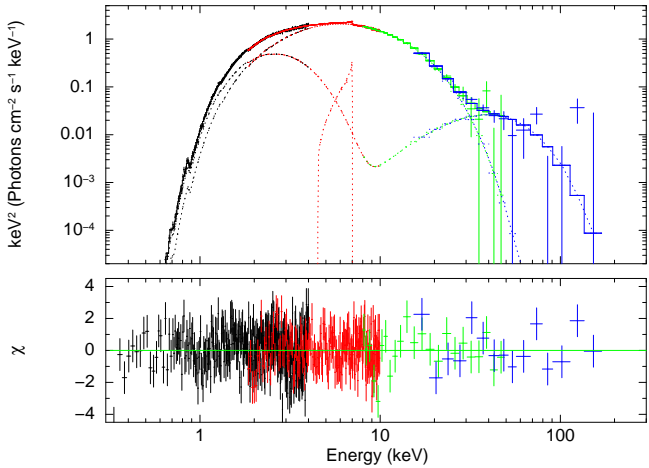}}
\vspace{0cm}
\parbox[b]{8.7cm}{\caption[]{Upper panel: unfolded averaged energy spectrum with model \textbf{Model4} and
single components in region B. Botton panel: residuals.}
\label{model4_s}}
\end{figure}

\begin{figure}[ht]
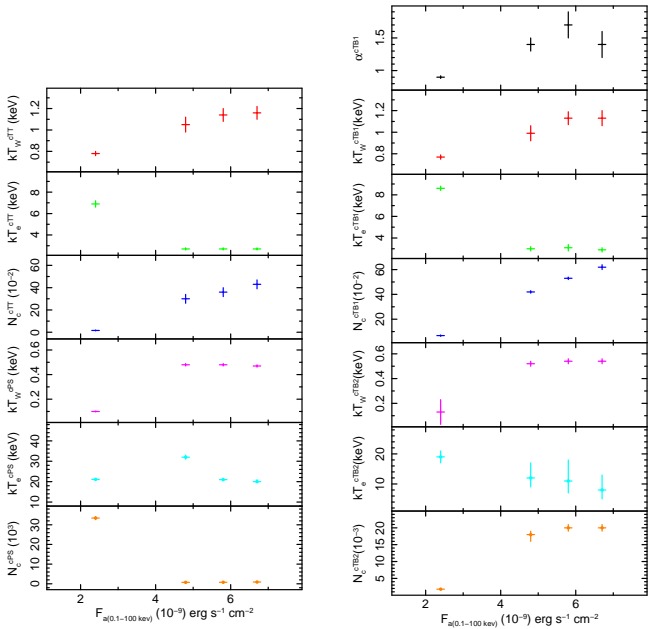
 
\begin{minipage}[t]{0.1\linewidth}
\vspace{2.1cm}
\vbox{ \psfig{file=1705_CPS_noref_2016.ps,height=4.0cm,angle=270.0}}
\end{minipage}
\hspace{3.5cm}
\begin{minipage}[t]{0.1\linewidth}
\vspace{1cm}
\vbox{ \psfig{file=1705_2tb_2016.ps,height=4.0cm,angle=270.0}}

\end{minipage}

\parbox[b]{8.7cm}{\caption[]{Best-fit parameters of the Comptonization components vs the absorbed flux in the range 0.1-100 keV  obtained using two different double Comptonization model. Left panel: values for \textbf{Model3}. Right panel:values for \textbf{Model4}. }\label{parameters2}}
\end{figure}

\begin{figure}[] 
\centering
\vspace{0.5cm}
%\vbox{\psfig{file=figure_2015/luglio2015/alpha1_Nctb1.ps,height=8.5cm,angle=270.0}}
\vbox{\psfig{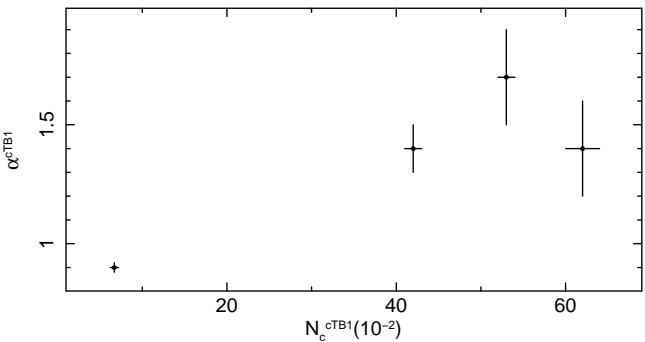}}
\vspace{0cm}
\parbox[b]{8.7cm}{\caption[]{Spectral index alpha  versus normalization of  the first {\tt CompTB}  component.}
\label{alpha1}}
\end{figure}

Since our aim is to model the two states with the same model components,  we tested double Comptonization models with nonreflection to highlight the evolution of the parameters. The first, \textbf{Model3}, is obtained by turning off the reflection in the {\tt compPS} component of \textbf{Model2} and the second, \textbf{Model4} consists in a double  {\tt compTB } model. The best-fit parameters of the two models are reported in Table 2. Following Titarchuk et al. 2014 in the framework of the applied model  {\tt{wabs*(compTB1+compTB2+diskline)}} the free
parameters of the model are $\alpha_{\rm 1/2}$,  $kT_{\rm c1/2}^{\rm i}$,  $log(A_{\rm 1/2})$, related to the
Comptonized fractions $f_{\rm 1/2}$,  and $kT_{\rm c1/2}^{\rm e}$, $N_{\rm C1/2}$, which are the normalizations of the  {\tt bb} components of the {\tt compTB1} und  {\tt compTB2}. We fixed certain parameters of the {\tt compTB } models, $\gamma_{\rm 1/2}$= 3 (related to the index of the low energy part of the spectrum, namely $\alpha_{\rm 1/2}$ =$\gamma_{\rm 1/2} -1$=2)  and $\delta_{\rm 1/2}$ = 0, because we neglect the bulk inflow effect versus the thermal Comptonization for neutron star accretion. The bulk inflow Comptonization takes place very close to neutron star surface. However, if the radiation pressure in that vicinity is sufficiently high then the bulk motion is suppressed. On the other hand, if mass accretion is low then the effect of the bulk motion is negligible. 
We also fixed the value of the  {\tt compTB1}  parameter $log(A_{\rm 1})$ to 8 when the best-fit values of   $log(A_{\rm 1})>>1$ because in any case of  $log(A_{\rm 1})>>1$ a Comptonization fraction $f = A/(1 + A)$ approaches unity and further variations of $A >> 1$ do not improve fit quality any more (Titarchuk et al. 2014). Finally, as the values of the  parameters $\alpha_{\rm 2} $ in the banana spectra B1, B2, and B3 are very small, we fixed them to $0$.
We did not introduce an extra {\tt{bb}}  component in \textbf{Model4}, which is required in the banana spectra, because it is modeled inside the {\tt{compTB2}} component.
In Fig.~\ref{model4_h} and \ref{model4_s} we show best-fit components and residuals  of  the  \textbf{Model4} applied  on I  and  the averaged B spectra, respectively.
Using \textbf{Model3}   and \textbf{Model4}, we investigated the spectral behavior versus the position on the CC diagram using the 0.1-100 keV flux  to identify the four regions. As shown in Fig.~\ref{parameters2}, the parameters of the two Comptonization components  vary significantly with the position along the CC track. 
We also studied the relation between the spectral index $\alpha_{\rm 1}$ of the first {\tt compTB} and its normalization $N_{\rm c1}$. This is shown in Fig.~\ref{alpha1}. 

\section{Discussion}

\subsection{The island state}

The Comptonization model of Poutanen \& Svensson (1966), which takes into account that a fraction of
Compton-produced photons can be reflected by cold matter (disk), successfully described the spectrum of the island state, that is when a spherical geometry of the Compton cloud is assumed together with reflection from a cold disk with solar abundances. 
More specifically, the model suggests that photons of temperature of $kT_i \simeq0.7$~keV are Comptonized in a region of electron temperature
$kT_e\sim27$~keV, which is characterized by $R_{i}$= 11~km. This scenario is naturally interpreted in the context of a truncated accretion disk scenario with the seed input photons coming from the region between the neutron star  surface and  inner part of the truncated accretion disk, or the boundary layer.  
The addition of a blackbody component to this model was not statistically significant, suggesting that we did not observe the seed photon component. 
Recently, Egron et al. (2013), combining data from XMM-Newton, RXTE, and BeppoSAX (with the exclusion of the PDS instrument), obtained a similar spectral deconvolution using the model  {\tt phabs*(nthComp+highecut*rdblur*reflionx)}, which contains a  thermally Comptonized continuum  ({\tt nthComp}; Zdziarski et al. 1996, Zycky et al. 1999) plus a self-consistent reflection model including both the reflection continuum and the corresponding discrete features (Rossi \&  Fabian 2005). The relativistically  smeared {\tt rdblur}  kernel (Fabian et al., 1989) was also used. The best-fit parameters that we obtained using the  model above are consistent with those reported  by Egron et al. (2013). 
We therefore confirm that during the hard state (observed in October 2000), a viable scenario is given by an absorbed  Comptonized component with evidence of reflection from a colder accretion disk (Fabian et al. 1989).
The presence of a broad iron emission line at about 6.4~keV and a Compton hump in the 20-40~keV energy range is expected from the  reprocessing of the main Comptonization continuum on the accretion disk. 
However,  to find a scenario that could be used in both the island and the banana state, we  explored an alternative 
description of the broadband energy spectrum of the low/hard state of  the source: a dual Comptonization of photons in two
distinct emission regions. 
We used {\tt compTT} plus {\tt compPS}, or two {\tt compTB}. The dominant emission results, by the Comptonization, in a disk geometry of seed photons at $\sim$0.1~keV by a plasma with kT$_{e}$ $\sim$20~keV and
optical depth $\sim$3. This emission can be interpreted as arising in a region above the accretion disk and contributes to the 70$\%$
of the total flux. We call this component hard Comptonization.
In addition, a secondary emission component, described by a spherical geometry, is due to seed photons at $\sim$0.8~keV Comptonized by a plasma with kT$_{e}$ $\sim$7~keV and optical depth
$\sim$8 using a {\tt compTT} model or with kT$_{e}$ $\sim$9~keV using a {\tt compTB}. This emission can be interpreted as originating from a region closer to the neutron star around the boundary layer and contributes about 30$\%$ of the total flux. We call this component soft Comptonization.
The persistent emission spectrum of the LMXB GS~1826-23, observed jointly by
Chandra and RXTE, has been described by Thompson et~al.\ (2005) with a similar model. This model uses dual Comptonization of $\sim$0.3~keV soft photons by a plasma with kT$_e$ $\sim$20~keV and $\tau\sim$2.6, which is interpreted as
emission from accretion disk corona, plus Comptonization of hotter 0.8 keV
photons by a $\sim$6.8~keV plasma, which is interpreted as emission from the
boundary layer. The parameters are therefore very similar to those reported here for 4U~1705-44.  
Moreover, a dual Comptonization model was successfully applied to the broadband spectra 
of the low/hard state of the black hole candidate Cyg X-1 by Makishima et ~al.\  (2008) with Suzaku data, by Di Salvo et ~al.\ 2001b  and Frontera et ~al.\  2001 with BeppoSAX data, and by Gierliski  et ~al.\  2007 and Ibragimov et ~al.\ 2005, who used simultaneous Ginga, RXTE and GRO/OSSE data. 
Takahashi et  ~al.\  2008 modeled  the spectra of GRO J1655--40 taken with Suzaku in a similar way. 
Recently, the two {\tt compTB} model was applied to six neutron star LMXBs (see Seifina et al. 2015). This model describes a scenario in which the Keplerian disk is connected to the neutron star by the transition layer (see Titarchuk et al. 1998), where the hot electrons scatter off  the soft photons from the neutron star photosphere of temperature kT$_1$ and the soft disk photons of temperature kT$_2$ giving rise to two Comptonized components. We stress that no reflection component is needed in the two scenarios described above. 

\subsection{Banana states and high energy tail}

The absorbed Comptonized component with reflection does not describe the soft/high state spectrum. In addition to
a dominating Comptonized emission, the continuum broadband spectrum of the source during the soft state requires  a blackbody component at lower energies and a hard component to model the hard tail or excess.
The dual Comptonization scenario successfully describes the broadband spectra of the banana state in that it models the 
soft dominating  emission and the hard tail well. Contrary to  the island state, the dominating Comptonization emission is soft   and is expected to originate in the hot plasma surrounding the neutron star.
The parameters of this component  change significantly when the source moves in the  banana branch and with respect to the island state.
Its contribution to the flux  changes from 57\% to 66\% for the three banana states.

The hard Comptonization shapes the hard excess emission originating from the disk region and its contribution to the flux is $\sim$17\% of the 0.1--200~keV source flux and changes with the CCD position. 
With respect to the island state flux, the flux of the soft Comptonization  is increased  by 400\%. The flux of the hard Comptonization  is reduced by 25\%.  
The low energy emission is modeled by a blackbody that we interpret as originating from the accretion disk. The
blackbody component contributes $\sim$22\% of the observed 0.1--200~keV flux and changes with the position in the CCD.

We note that a different scenario was analyzed in Piraino et al. (2007). 
In that case, we described the average banana branch spectrum by the sum of a blackbody, a Comptonized component, and a hard energy tail.
The hard energy tail, described with a steep power-law with photon index $\sim$2-3 without evidence of a
high energy cutoff, could be produced, as proposed for Z-sources, either in a hybrid thermal/nonthermal model (Poutanen \& Coppi 1998) or in a bulk motion of matter close to the neutron star (e.g. Titarchuk \&
Zannias 1998). The alternative mechanism proposed is the Comptonization of seed
photons by high-velocity electron of a jet (e.g. Di Salvo et~al.\ 2000) or the
synchrotron emission from a relativistic jet escaping the system (Markoff,
Falcke \& Fender 2001). We refer to Piraino et al. (2007) for details on this scenario and on similarities with black holes.  

\section{Conclusions}

In this Note, we have presented a spectral-timing analysis of the full set of the BeppoSAX observations of \4U. 
In our analysis, for the first time ever, all available data from the BeppoSAX NFIs were used. 
As a result of our analysis, we obtained 1) the broadband X--ray spectra of the source, from soft to hard X-rays (0.3--200~keV) for different spectral states, that is for the island and banana branches;
2) the PSDs in the island and banana branches and the CCD diagram for all available observations; and 3) broadband spectra resolved with respect to the position of the source in the CCD diagram.

Using these results, we confirmed through the PSD that the source was, indeed, in hard state during OBS2 and we succeeded in modeling the spectra of the source using the same physical scenario for both the island and banana branches (the double Comptonization scenario). 

According to this scenario, the following components 
are observed: 
1) soft  Comptonization arising from the hot plasma surrounding the neutron star and dominant during the banana state;
2) hard Comptonization from the disk region and dominant during the island state;
3) blackbody component arising from the inner part of the accretion disk and not observed in the island state; 
and 4) broad features in the $K_\alpha$ iron line region observed in both banana and island states, which are modeled well  with the {\tt diskline}  model.

The evidence  of a possible reflection component in the broadband energy spectrum of the island state favors the interpretation of the iron line as a emission from the accretion disk. However, in the double Comptonization scenario no reflection is required. We cannot therefore exclude that broad lines can originate in the outflow from the disk. 

Finally, we wish to stress that the double Comptonization model appears as the more suitable solution to describe the spectral behavior of the source during all different  states observed with BeppoSAX.

\begin{acknowledgements}
This work was  supported by INAF and IAAT. The High-Energy Astrophysics Group of Palermo acknowledges support by
the Fondo Finalizzato alla Ricerca (FFR) 2012/13, project N. 2012-ATE-0390,
founded by the University of Palermo, by the Region of  Sardinia
through POR-FSE Sardegna 2007-2013, L.R. 7/2007, Progetti di
Ricerca di Base e Orientata, Project N. CRP-60529, and by the INAF/PRIN 2012-6. E. Egron acknowledges financial support from the Autonomous Region of Sardinia through a research grant under the program CRP-25399 PO Sardegna FSE 2007-2013, L.R. 7/2007, promoting scientific research and innovation technology in Sardinia. We thank the anonymous referee for her/his comments and Amy Mednick for language editing.
\end{acknowledgements}

\end{document}